\documentclass[review,number,sort&compress,longtitle]{elsarticle}
\usepackage{graphicx,amsmath,amsfonts,amscd,amsthm,amssymb,pstricks,lineno}
\makeatletter
\newcommand*{\Hy@backout}[1]{}
\makeatother

\newcommand{\dif}[1]{\ensuremath{\mathrm{d}#1}}
\newcommand{\Int}[3]{\ensuremath{\int\limits_{#1}^{#2}\!\dif{#3}\;}}
\newcommand{\Intnl}[3]{\ensuremath{\int_{#1}^{#2}\!\dif{#3}\;}}
\newcommand{\Oint}[3]{\ensuremath{\oint\limits_{#1}^{#2}\!\dif{#3}\;}}
\newcommand{\Exp}[1]{\ensuremath{\mathrm{e}^{#1}}}
\begin{document}
\begin{frontmatter}
\title{Description of the luminosity evolution for the CERN LHC including dynamic aperture effects. Part I: the model\tnoteref{mytitlenote}}
\tnotetext[mytitlenote]{Research supported by the HL-LHC project}
\author[cern]{M. Giovannozzi\corref{mycorrespondingauthor}}
\cortext[mycorrespondingauthor]{Corresponding author}
\ead{massimo.giovannozzi@cern.ch}
\author[cern]{F.F.~Van~der~Veken}
\address[cern]{Beams Department, CERN, CH 1211 Geneva 23, Switzerland}
\begin{abstract}
In recent years, modelling the evolution of beam losses in circular proton machines starting from the evolution of the dynamic aperture has been the focus of intense research. Results from single-particle, non-linear beam dynamics have been used to build simple models that proved to be in good agreement with beam measurements. These results have been generalised, thus  opening the possibility to describe also the luminosity evolution in a circular hadron collider. In this paper, the focus is on the derivation of scaling laws for luminosity, which include both burn off and additional pseudo-diffusive effects. It is worthwhile stressing that time-dependence of some beam parameters can be taken into account in the proposed framework. The proposed models are applied to the analysis of a subset of the data collected during the CERN Large Hadron Collider (LHC) Run~1 in a companion paper~\cite{lumi_Part_II}.
\end{abstract}
\end{frontmatter}
%
%\linenumbers
% 
\section{Introduction}
Since the advent of the generation of superconducting colliders, the unavoidable non-linear magnetic field errors have plagued the dynamics of charged particles inducing new and potential harmful effects. This required the development of new approaches to perform more powerful analyses and to gain insight in the beam dynamics. Parenthetically, in a number of cases, rather than developing new tools, accelerator physics imported existing tools from other domains of science, such as celestial mechanics, dealing with similar problems. Among the several studies that flourished in the nineties, that on the concept of dynamic aperture (DA), i.e. the region in phase space where bounded motions occur, was particularly fruitful. The literature on this topic is very large, and most of it beyond the scope of the results presented in this paper. Nevertheless, it is worth mentioning the work done on the scaling law of the DA as a function of time~\cite{dynap1,dynap2} for the case of single-particle beam dynamics. Indeed, such a scaling law was later successfully extended to the case in which weak-strong beam-beam effects are added to the beam dynamics~\cite{loginvb-b}. More importantly, such a scaling law was proposed to describe the time evolution of beam losses in a circular particle accelerator under the influence of non-linear effects~\cite{lossesPRSTAB}, and the proposed model was verified experimentally, using data from CERN accelerators and the Tevatron. Note that such a scaling law for beam intensity as a function of time is at the heart of a novel method to measure experimentally the DA in a circular ring~\cite{DAexp_nekor}.

The model developed represents a bridge between the concept of DA, which is rather abstract, and the beam losses observed in a particle accelerator. Clearly, the next step was to extend the model to describe the luminosity evolution in a circular collider. The first attempts are reported in~\cite{Lumi_fit,IPAC14}. However, in those papers the DA scaling law was used without disentangling the contribution of burn off. Although the results were rather encouraging, to recover the correct physical meaning of the model parameters it was necessary to include as many known effects as possible.

This limitation is removed in the model discussed in this paper. In fact, the proposed scaling law is combined with the well-know intensity decay from particles' burn off so that a coherent description of the physical process is provided. Moreover, it is worth stressing that the proposed model can be generalised so to consider a time-dependence for some of the beam parameters describing the luminosity evolution, such as emittance. 

The plan of the paper is the following: in section~\ref{sec:model} the general model of luminosity evolution is presented, starting from the case with burn off and time-independent beam parameters (section~\ref{sec:general}). In section~\ref{sec:new_model} the proposed model is presented, which includes also pseudo-diffusive contributions from non-linear effects. This model is then shown in action in section~\ref{sec:int_lumi}, where these concepts are applied to the problem of the analysis of the distribution of the integrated luminosity whereas in section~\ref{sec:optfill} the analysis of the optimal physics fill duration is dealt with. The implications of using different models, namely plain burn off or burn off together with pseudo-diffusive phenomena are also addressed. Conclusions are drawn in section~\ref{sec:conclusions} while additional detail is presented in the Appendices. Additional refinements to the burn-off model aimed at including time-dependent emittance variations are discussed in \ref{app:refinement}, while computations for the case of plain burn off in conjunction with explicit time dependence of some of the beam parameters are reported in \ref{app:time-dip}. The detail of the derivation of the proposed model is worked out in \ref{app:derivation}.
\section{Luminosity evolution with proton burn off losses\label{sec:model}} 
\subsection{General considerations}\label{sec:general}
The starting point is the expression of luminosity, which is a key figure-of-merit for colliders and, neglecting the hourglass effect, reads
\begin{equation}
L = \frac{\gamma_{\rm r} \, f_{\rm rev} \, k_{\rm b} \, n_1 \, n_2}%
{4 \, \pi \epsilon^* \beta^*} \, F(\theta_{\rm c}, \sigma_{z}, \sigma^* ),
\label{lumi}
\end{equation}
where $\gamma_{\rm r}$ is the relativistic $\gamma$-factor, $f_{\rm rev}$ the revolution frequency, $k_{\rm b}$ the number of colliding bunches, $n_{\rm i}$ the number of particles per bunch in each colliding beam, $\epsilon^*$ is the RMS normalised transverse emittance, and $\beta^*$ is the value of the beta-function at the collision point. The total beam population is defined as $N_{j}=k_{\rm b} \, n_{j}$ and the fact that not all bunches are colliding in the high-luminosity experimental points is taken into account by introducing a scale factor.

The factor $F$ accounts for the reduction in volume overlap between the colliding bunches due to the presence of a crossing angle and is a function of the half crossing angle $\theta_{\rm c}$ and the transverse and longitudinal RMS dimensions $\sigma^*, \sigma_{z}$, respectively according to:
\begin{equation}
F(\theta_{\rm c}, \sigma_{z}, \sigma^* )=\frac{1}{\sqrt{1+
\left ( \displaystyle{\frac{\theta_{\rm c}}{2} \, \frac{\sigma_{z}}{\sigma^*}} 
\right )^2}} \, .
\label{geofac}
\end{equation}
Note that $\sigma^*=\sqrt{\beta^* \, \epsilon^*/(\beta_{\rm r} \, \gamma_{\rm r})}$, where $\beta_{\rm r}$ is the relativistic $\beta$-factor. Equation~(\ref{lumi}) is valid in the case of round beams ($\epsilon_x^*=\epsilon_y^*=\epsilon^*$) and round optics ($\beta_x^* = \beta_y^*=\beta^*$). For our scope, Eq.~(\ref{lumi}) will be recast in the following form:
\begin{equation}\label{LumiDef}
L = \Xi \, N_1 \, N_2,  \qquad  \Xi = \frac{\gamma_{\rm r} f_{\rm rev}}%
{4 \, \pi \epsilon^* \beta^* \, k_{\rm b} } 
F(\theta_{\rm c}, \sigma_{z}, \sigma^* )
\end{equation}
in which the dependence on the total intensity of the colliding beams is highlighted and the other quantities are included in the term $\Xi$ .

Under normal conditions, i.e. excluding any levelling gymnastics or dynamic-beta effects, only the emittances and the bunch intensities can change over time. Therefore, Eq.~(\ref{lumi}) is more correctly interpreted as peak luminosity at the beginning of the fill, as in general $L$ is a function of time. When the burn off is the only relevant mechanism for a time-variation of the beam parameters, it is possible to estimate the time evolution of the luminosity, which turns out to be derived from the following equation
\begin{equation}
{N}'(t) = - \sigma_{\rm int} \, n_{\rm c} \, L(t) = - 
\sigma_{\rm int} \, n_{\rm c} \, \Xi \, N^2(t)
\end{equation}
where $\sigma_{\rm int}$ represents the cross section for the interaction of  charged particles and the two colliding beams have been assumed to be of equal intensity. The value used is $73.5$~mb for 3.5~TeV and $76$~mb for 4~TeV~\cite{inel1,inel2} for protons, representing the total inelastic cross-section. Here, $n_{\rm c}$ stands for the number of collision points. The solution is given by
\begin{equation}
N(t)= \frac{N_{\rm i}}{1+\sigma_{\rm int} \, n_{\rm c} \, \Xi \, N_{\rm i} \, t},
\label{eq:int_evol}
\end{equation}
where $N_{\rm i}$ stands for the initial intensity. Equation~(\ref{eq:int_evol}) implies that the luminosity evolves as 
\begin{equation}
L(t) = 
\frac{\Xi \, N^2_{\rm i}}%
{\left ( 1 + \sigma_{\rm int} \, n_{\rm c} \, \Xi \, N_{\rm i} \, t \right )^2} \,.
\label{burnoff}
\end{equation}

In the most general case, where both beams can have different intensities, the intensity evolution is described by the following equations
\begin{equation}
\begin{cases}
{N}'_1(t) & = -\sigma_{\rm int} \, n_{\rm c} \, \Xi \, N_1(t) \, N_2(t)  \\
{N}'_2(t) & = -\sigma_{\rm int} \, n_{\rm c} \, \Xi \, N_1(t) \, N_2(t)  \, . 
\end{cases}
\label{int_main}
\end{equation}
For reasons that will become clear later, a different time variable is considered, namely
\begin{equation}
\tau -1 = f_{\rm rev} \, t \qquad \text{giving} \qquad \frac{\dif}{\dif t} = 
f_{\rm rev} \, \frac{\dif}{\dif \tau} \, ,
\end{equation}
$\tau$ being an adimensional variable representing the number of turns, where a shift of the origin of $\tau$ with respect to $t$ has been introduced. In the following, the derivative with respect to $\tau$ is indicated by $\dot{\phantom{o}}$, while $'$ indicates the derivative with respect to $t$. 

The solution of Eq.~(\ref{int_main}), indicated as $N^{\rm bo}_{1,2} (\tau)$ to highlight that it only includes the burn off contribution, can be obtained by re-writing:
\begin{equation}
\begin{cases}
\dot{N}^{\rm bo}_1(\tau) + \dot{N}^{\rm bo}_2(\tau) & = -2 \, \varepsilon \, 
N^{\rm bo}_1(\tau) \, N^{\rm bo}_2(\tau) \\
\dot{N}^{\rm bo}_1(\tau) - \dot{N}^{\rm bo}_2(\tau) & = 0 
\end{cases}
\label{prototype}
\end{equation}
with
\begin{equation}\label{eq:epsilon}
\varepsilon=
 \frac{
 	\sigma_{\rm int} \, n_{\rm c} \, \Xi
    }{f_{\rm rev}}
\end{equation}
and from which one finds
\begin{equation}
\begin{cases}
N^{\rm bo}_1(\tau)  & = N^{\rm bo}_2(\tau) + \xi \\
\dot{N}^{\rm bo}_2(\tau) & = - \, \varepsilon \, N^{\rm bo}_2(\tau) \, 
\left [ N^{\rm bo}_2(\tau) + \xi \right ] . 
\end{cases}
\label{int_main1}
\end{equation}
Equation~(\ref{int_main1}) has two solutions depending on the value of $\xi$. If $\xi=0$ then
\begin{equation}
\begin{cases}
N^{\rm bo}_1(\tau) & = \displaystyle{\frac{N_{\rm i}}{1 + \varepsilon \, N_{\rm i} 
\, (\tau-1)}} \\
N^{\rm bo}_2(\tau)  & = N^{\rm bo}_1(\tau),
\end{cases}
\label{ssol1}
\end{equation}
where $N_{\rm i}=N_{{\rm i},1}=N_{{\rm i},2}$ stands for the initial beam intensity. 

Otherwise, if $\xi \neq 0$ then
\begin{equation}
\begin{cases}
N^{\rm bo}_1(\tau) & =
	\xi \; \displaystyle
	\frac{1} {1- N_{\rm r} \,\Exp{-\varepsilon \,\xi\, (\tau -1)}} \\[1.5em]
N^{\rm bo}_2(\tau) & =
	\xi \; \displaystyle
	\frac{N_{\rm r} \; \Exp{-\varepsilon \,\xi\, (\tau -1)}}
					{1- N_{\rm r} \,\Exp{-\varepsilon \,\xi\, (\tau -1)}} \, ,
\end{cases}
\label{ssol2}
\end{equation}
where $\xi=N_{{\rm i},1} - N_{{\rm i},2}$ and $N_{\rm r} = \frac{N_{{\rm i},2}}{N_{{\rm i},1}}$. Note that the result does not depend on the sign of $\xi$, although, from a computational point of view a negative exponent is preferred, as $\tau$ can grow to very large values. It is possible to enforce this by redefining $\xi=N_{\rm i,max} \!-\! N_{\rm i,min}$ and $N_{\rm r} = \frac{N_{\rm i,min}}{N_{\rm i,max}}$, with $N_{\rm i,min}=\min{\left\{N_{i, 1},N_{i, 2}\right \}}$ and $N_{\rm i,max}=\max{\left\{N_{i, 1},N_{i, 2}\right\}}$. This gives the same result as in Eq.~\eqref{ssol2} using $N_{\rm i,max}$ and $N_{\rm i,min}$ instead of $N_{{\rm i},1}$ and $N_{{\rm i},2}\,$. Note also that Eq.~\eqref{ssol1} can be easily recovered from Eq.~\eqref{ssol2} by expanding the exponential and carefully setting $N_{{\rm i},1}=N_{{\rm i},2}$.

\begin{figure}[htb]
  \begin{center}
    \begin{tabular}{@{}cc@{}}
      \includegraphics[width=0.48\linewidth,clip=]{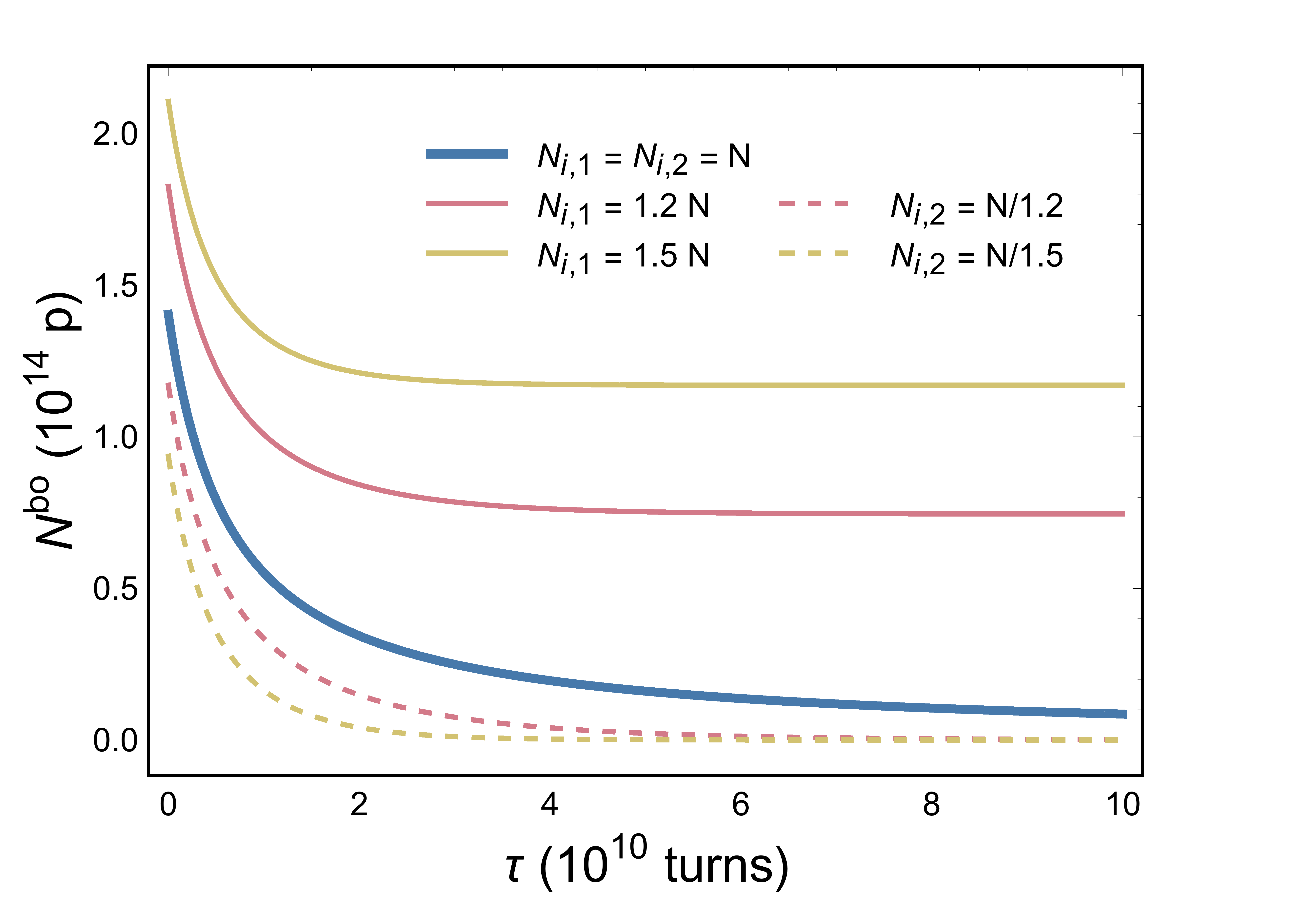} &
      \includegraphics[width=0.48\linewidth,clip=]{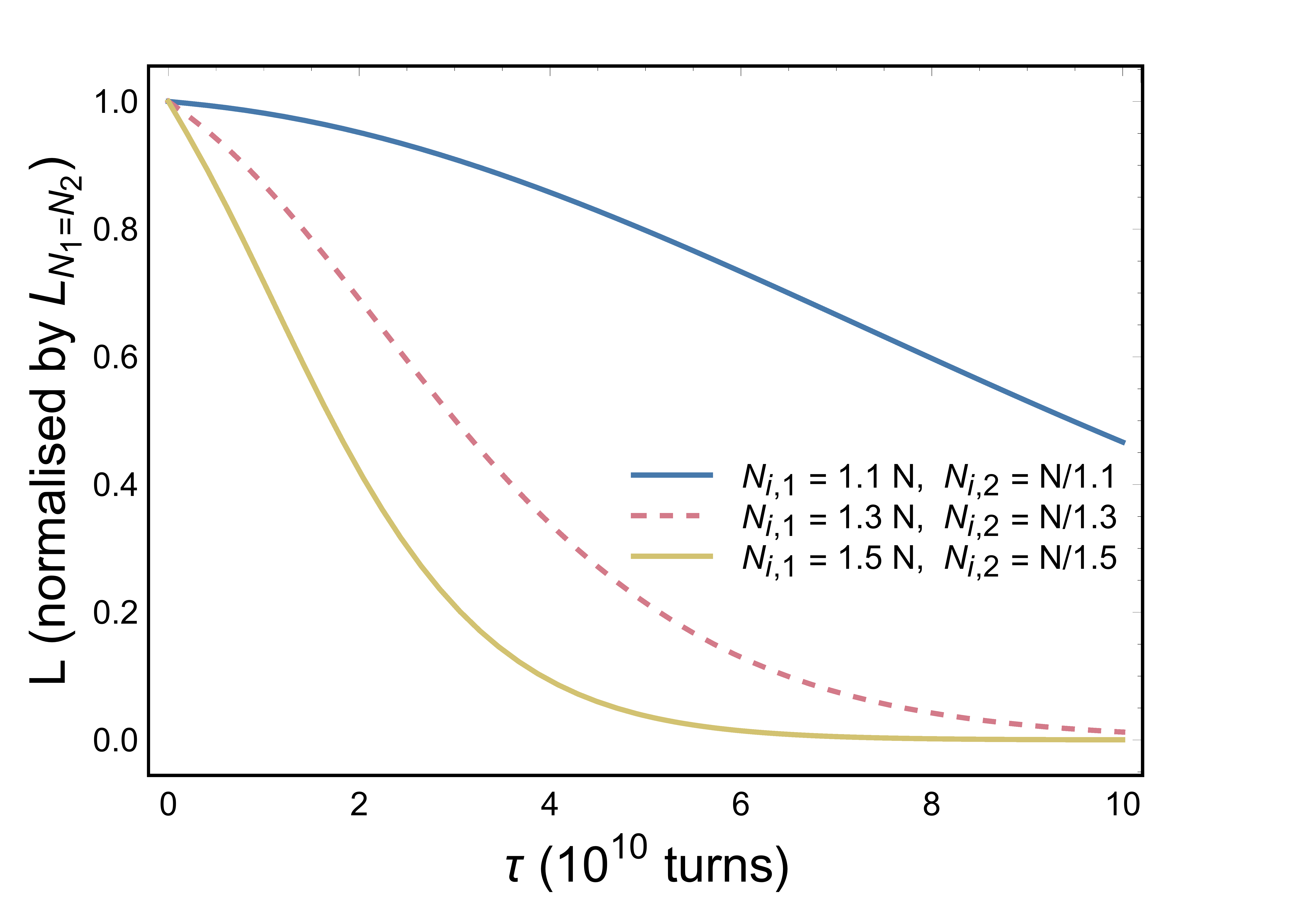}
    \end{tabular}
  \end{center}
  \caption{Evolution of intensity (left) and luminosity (right) based on the solutions of Eqs.~(\ref{ssol1}) and~(\ref{ssol2}), respectively. For the plots presented here, $\varepsilon=1.1\times 10^{-24}, n_{\rm i} =1\times 10^{11}, k_{\rm b}= 1404$, which are values representative for the proton beam parameters during the Run~1 of the LHC. The luminosity evolution is normalised to the value corresponding to the solution with equal intensities. The different behaviour is clearly visible. The case with equal intensities turns out to be the best scenario as far as the luminosity is concerned.}
  \label{example_evol}
\end{figure}

As an example, in Fig.~\ref{example_evol} the behaviour of the intensity and luminosity evolution is shown for various cases with equal or different intensities for the two beams. 

To compare the results, the constraint $N_{\rm i, 1} N_{\rm i, 2} = N^2$ has been imposed. The differences are clearly visible. Firstly, the time behaviour is exponential-like for the case of different intensities, thus generating shorter fills. Secondly, in case of unequal intensity one beam is completely burnt while the other is not, which makes the use of the protons in the collider inefficient. Furthermore, the evolution of the relative luminosity, i.e. the luminosity normalised to the corresponding value of the case with equal beam intensities, shows that the most favourable situation is the one featuring equal beam intensities, as any imbalance in beam intensity worsen the collider performance. This is due to the fact that the burn off is the same for both beams, hence the maximum fill length is given by the time needed to burn the weaker beam. 

Whenever additional time dependence in the luminosity evolution needs to be taken into account,  and assuming that a model for such a time-dependence is available~\cite{fnal1,fnal2}, then the solutions~(\ref{ssol1}) and~(\ref{ssol2}) can be extended to take into account these effects. The detail is presented in \ref{app:refinement} and \ref{app:time-dip}.
\section{Luminosity evolution including pseudo-diffusive effects}\label{sec:new_model}
An efficient modelling of the luminosity evolution in a real collider can be obtained either by means of numerical tracking or by means of analytical or semi-analytical models, grasping the main features of the luminosity evolution. The first approach has been used, e.g. in Ref.~\cite{bruce}, where the luminosity evolution is studied by means of direct numerical simulations taking into account all relevant physical processes. The second approach, has been used in Refs.~\cite{fnal1,fnal2}, where phenomenological fit models were proposed and applied with success to the characterisation of luminosity evolution in the Tevatron machine. The functional form of the proposed models was suggested by considerations on scaling laws of key quantities such as emittances. Another example is given in Ref.~\cite{FZ}, where a refined semi-analytical model including several phenomena affecting emittance variation has been derived in view of optimising the performance of future circular colliders. However, none of the models studied included the effect of non-linear motion and this is at the heart of the approach proposed in Ref.~\cite{Lumi_fit}. The basis for such a model is the evolution of the dynamic aperture (DA) with time in a hadron collider. The analysis of single-particle tracking results showed that the time evolution of the DA follows a simple law~\cite{dynap1,dynap2}, whose justification is not entirely phenomenological. Recently, this approach has been successfully applied to the analysis of intensity evolution in hadron machines~\cite{lossesPRSTAB}. So far, however, the results were obtained in the case of single-particle simulations or whenever the conditions in a particle accelerator were not under the influence of any collective effect. To extend the proposed scaling law to luminosity evolution, it is necessary to show that it is valid also in the presence of beam-beam effects. This is the case at least for the results of numerical simulations in the weak-strong regime, as discussed in Ref.~\cite{loginvb-b}, thus opening the possibility to justify the proposed interpretation. 

The proposed approach is a refinement of what is presented in Ref.~\cite{Lumi_fit} and assumes that all possible pseudo-diffusive effects can be modelled by a scaling of the intensity with time as
\begin{equation}
N(\tau)\, = N_{\rm i} \, \left [ 1 - \Int{D(\tau)}{+\infty}{r} \hat{\rho}(r)
\right ] =  N_{\rm i} 
\left [ 1 - \Exp{-\frac{D^2(\tau)}{2}} \right ] \, ,
\label{eqmain1}
\end{equation}
where
\begin{equation}  
D(\tau) = D_{\infty}  + \frac{b}{ \left [ \log \tau \right ]^{\kappa}}.
\label{main}
\end{equation}
The parameters $D_{\infty}, b, \kappa$ are normally fitted to the experimental data and the variable $\tau$ represents the turn number and satisfies $\tau \in [1, +\infty [$. It is worthwhile stressing some properties of the parameters as highlighted in Refs.~\cite{dynap2,lossesPRSTAB}, where two regimes were identified:
\begin{itemize}
\item in 4D systems the three quantities $D_{\infty}, b, \kappa$ are all positive~\cite{dynap1, dynap2}. This corresponds to having a stable region in phase space for an  arbitrarily long time and hence being in the condition of applicability of the Kolmogorov-Arnold-Moser (KAM) and Nekhoroshev theorems.
\item in 4D systems with tune modulation or 6D models, it is possible that there is no stable region even for a finite number of turns~\cite{dynap2}. This corresponds to one of the two cases:
\begin{equation}
\begin{cases}
 D_{\infty} & > 0 \qquad \qquad \kappa < 0 \qquad \qquad b < 0 \\
 D_{\infty} & \leq 0 \qquad \qquad \kappa > 0 \qquad \qquad b >0
 \end{cases}
\label{zero}
\end{equation}
The first case represents a situation with global chaoticity~\cite{dynap2}. In particular, the fact that tracking data could be fitted with negative $\kappa$ has been already observed several years ago (see discussion in Ref.~\cite{dynap2}). The latter is compatible with a situation in which the stable KAM area shrinks to zero and the escape to infinity is governed by a Nekhoroshev-like behaviour. Clearly, the possibility of having $D_{\infty} < 0 $ goes beyond the proposed picture based on phase space partitioning into regions in which the beam dynamics is governed by KAM and Nekhoroshev theorems.
\end{itemize}

The further step in view of using this scaling law for the analysis of the evolution of the luminosity requires a number of additional considerations, namely
\begin{itemize}
\item The proton burn off occurs mainly in the core of the beam distribution, corresponding to the region of largest particle density. On the other hand, the diffusive processes are mainly affecting the tails of the beam distribution. This, in turn, implies that proton burn off and diffusive phenomena are acting on different parts of the beam distributions and are, hence, essentially decoupled and independent.
\item The characteristic times of the two processes are rather different. The burn off takes place at a sub-turn time scale (for instance, in the case of the LHC, considering only the high-luminosity experiments, the burn off occurs twice per turn), while the pseudo-diffusive phenomena take place on a much longer time scale, as a continuous process.
\item The fit parameters in Eq.~(\ref{main}) might depend on the beam intensity. However, if one assumes that the overall intensity variation over one physics fill is not too large, it is then reasonable to consider that the pseudo-diffusive effects are, to a good extent, almost constant.
\end{itemize}
Then, under these assumptions, it is justified to describe the intensity evolution as
\begin{equation}
\begin{cases}
{N}'_1(t) & = -\sigma_{\rm int} \, n_{\rm c} \, \Xi \, N_1(t) \, N_2(t) - 
\hat{\mathcal{D}}_1(t) \\
{N}'_2(t) & = -\sigma_{\rm int} \, n_{\rm c} \, \Xi \, N_1(t) \, N_2(t) - 
\hat{\mathcal{D}}_2(t) \, . 
\end{cases}
\label{int_main0}
\end{equation}
The terms $\hat{\mathcal{D}}_i$ represent the intensity-independent pseudo-diffusive effects. The reason for a shift of origin of $\tau$ with respect to $t$ is now clear: it is needed so that $t=0$ corresponds to $\tau=1$ and then,  $D(1)=+\infty$ according to Eq.~(\ref{main}), which is the expected situation. 

Equation~(\ref{int_main}) becomes
\begin{equation}
\begin{cases}
\dot{N}_1(\tau) & = -\varepsilon \, N_1(\tau) \, N_2(\tau) - 
\mathcal{D}_1(\tau) \\
\dot{N}_2(\tau) & = -\varepsilon \, N_1(\tau) \, N_2(\tau) - 
\mathcal{D}_2(\tau) \, ,
\end{cases}
\label{int_main2}
\end{equation}
with $\mathcal{D}_i=\hat{\mathcal{D}}_i/f_{\rm rev}$. 
Typical values of $\varepsilon$ are $1.1\times 10^{-24}$ assuming the beam parameters during the 2011 physics run for protons. Therefore, about $3.1\times 10^4$ particles are removed from the bunches each turn, corresponding to $0.24$~ppb.

The explicit expression for $ \mathcal{D}_i(\tau)$ can be found by noting that these functions are the solutions of
\begin{equation}
\begin{cases}
\dot{N}_1(\tau) & = - \mathcal{D}_1(\tau) \\
\dot{N}_2(\tau) & = - \mathcal{D}_2(\tau) 
\end{cases}
%\label{int_main4}
\end{equation}
and that the explicit solution has been assumed to be of the form~(\ref{eqmain1})~\cite{Lumi_fit,lossesPRSTAB,loginvb-b}. Therefore, one obtains
\begin{equation}
\mathcal{D}_j(\tau)=-N_{{\rm i},j} \, D_j(\tau) \, \dot{D}_j(\tau) \, 
\Exp{-\frac{D^2_j(\tau)}{2}} \qquad j=1,2 \, .
\label{sol_main0}
\end{equation}

The detail of the derivation can be found in~\ref{app:derivation}, but under the assumptions that the initial beam intensities are the same as well as the terms $\mathcal{D}_{j}$, then an explicit expression at the lowest order in $\varepsilon$ (see Eq.~(\ref{eq:epsilon})) can be given for both intensity and luminosity, namely
\begin{equation}
\frac{N(\tau)}{ N_{\rm i}} = \displaystyle{\frac{1}{1 + \varepsilon \, N_{\rm i} 
\, (\tau-1)}- \left [ \Exp{-\frac{D^2(\tau)}{2}} - \Exp{-\frac{D^2(1)}{2}}\right ]}
\end{equation}
and
\begin{align}\label{Lpd unint}
\frac{L(\tau)}{L_{\rm i}} & = \displaystyle{\frac{1}{\left [ 1 + \varepsilon \, N_{\rm i} \, (\tau-1) \right ]^2} - 
\left [ \Exp{-\frac{D^2(\tau)}{2}} - \Exp{-\frac{D^2(1)}{2}}\right ]} \times \notag \\
& \phantom{=} \times \displaystyle{\left \{ 2- \left [ \Exp{-\frac{D^2(\tau)}{2}} - \Exp{-\frac{D^2(1)}{2}}\right ] \right \}}
\end{align}
where $L_{\rm i}$ is the initial value of the luminosity, given by $L_{\rm i}=\Xi \, N^2_{\rm i}$.

In case $\varepsilon=\varepsilon(\tau)$ is time-dependent, an analogous derivation can be applied as shown in \ref{app:time-dip} and \ref{app:derivation}. The result is very similar to the time-independent case, namely
\begin{equation}
\frac{N(\tau)}{ N_{\rm i}} = \displaystyle{\frac{1}{1 + N_{\rm i}\,\mu(\tau)}- \left [ \Exp{-\frac{D^2(\tau)}{2} \, \frac{\epsilon(1)}{\epsilon(\tau)}} - \Exp{-\frac{D^2(1)}{2}}\right ]}
\end{equation}
and
\begin{align}
\frac{L(\tau)}{L_{\rm i}} & = \displaystyle{\frac{1}{\left[1 + N_{\rm i}\,\mu(\tau) \right]^2} - 
\left [ \Exp{-\frac{D^2(\tau)}{2} \, \frac{\epsilon(1)}{\epsilon(\tau)}} - \Exp{-\frac{D^2(1)}{2}}\right ]} \times \notag \\
& \phantom{=} \times \displaystyle{\left \{ 2- \left [ \Exp{-\frac{D^2(\tau)}{2} \, \frac{\epsilon(1)}{\epsilon(\tau)}} - \Exp{-\frac{D^2(1)}{2}}\right ] \right \}} \,.
\end{align}
where $\mu(\tau)$ is defined in Eq.~\eqref{eq:mu}. Note that the time-independent and the time-dependent solutions are related by two substitutions, namely $\left(\tau-1\right)\,\varepsilon \to \mu(\tau)$ and $D^2(\tau) \to D^2(\tau) \,\epsilon(1)/\epsilon(\tau)$. The second substitution deserves some comments. The scaling law \eqref{eqmain1} implies that $D(\tau)$ is measured in terms of beam sigma, under the assumption that sigma does not change with time. In case of emittance growth, this assumption is violated and the proposed substitution simply takes into account the time-dependence of sigma (which scales with respect to the emittance as $\sigma \sim \sqrt{\epsilon}$). Note that this approach is possible because it assumes a complete independence of the pseudo-diffusive and emittance growth effects. This is fully justified by the assumption that the pseudo-diffusive effects are acting on the tails, while emittance growth affects the beam core. Note that these replacements have a general validity and will be true also for what follows.
\section{Integrated luminosity over a physics fill\label{sec:int_lumi}}
The models analysed in the previous sections can be used to derive some useful scaling laws for the integrated luminosity as a function of the length of the physics fill. Indeed, assuming the simple case of equal intensities for both beams, it is possible to obtain for the burn off part
\begin{equation}
\hat{L}_{\rm int}^\text{bo} (t) =
\Int{0}{t}{\tilde{t}}\hat{L}^\text{bo}(\tilde{t}) =
	\frac{N_{\rm i} \, \Xi}{\varepsilon \, f_{\rm rev}} \, 
	\frac{\varepsilon \, N_{\rm i} \, f_{\rm rev} \, t}
		{1+ \varepsilon \, N_{\rm i} \, f_{\rm rev} \, t}
\, ,
\end{equation}
which is obtained by means of Eq.~\eqref{burnoff} and the very definition of $\varepsilon$.
Note the limiting value 
\begin{equation}
\hat{L}_{\rm int}^\text{bo} (t \to \infty)= \frac{N_{\rm i} \, \Xi}{\varepsilon \, f_{\rm rev}} \, ,
\end{equation}
so that one can normalise the integrated luminosity as
\begin{equation}\label{lumint1}
\hat{L}_{\rm norm}^\text{bo}(t) =\frac{\hat{L}_{\rm int}^\text{bo}(t)}{\hat{L}_{\rm int}^\text{bo}(t \to \infty)} = 
	\frac{\varepsilon \, N_{\rm i} \, f_{\rm rev} \, t}
		{1+ \varepsilon \, N_{\rm i} \, f_{\rm rev} \, t}
\, .
\end{equation}
It is straightforward to express the integrated luminosity as a function of $\tau$, starting from Eqs.~\eqref{LumiDef} and \eqref{ssol1}, and integrating over $\left[1,\tau\right]$, namely
\begin{equation}
L_{\rm int}^\text{bo} (\tau) =
	\Int{1}{\tau}{\tilde{\tau}} L^\text{bo}(\tilde{\tau}) =
	\frac{N_{\rm i} \, \Xi}{\varepsilon} \, 
	\frac{\varepsilon \, N_{\rm i} \, (\tau-1)}{1+ \varepsilon \, N_{\rm i} \, (\tau-1)} \, .
\end{equation}
Hence $\hat{L}_{\rm int}^\text{bo}(t)$ and $L_{\rm int}^\text{bo}(\tau)$ are related by a scale factor $1/f_{\rm rev}\,$:
\begin{equation}
	\hat{L}_{\rm int}^\text{bo} (t) = \frac{1}{f_{\rm rev}} L_{\rm int}^\text{bo} (\tau) \, .
\end{equation}
Note that because
\begin{equation}
L_{\rm int}^\text{bo} (\tau \to \infty) =
	\frac{N_{\rm i} \, \Xi}{\varepsilon} \, ,
\end{equation}
$\hat{L}_{\rm norm}^\text{bo}(t)$ and $L_{\rm norm}^\text{bo}(\tau)$ are now equal:
\begin{equation}\label{luminttau}
L_{\rm norm}^\text{bo}(\tau) =\frac{L_{\rm int}^\text{bo}(\tau)}{L_{\rm int}^\text{bo}(\tau \to \infty)} = 
	\frac{\varepsilon \, N_{\rm i} \, (\tau-1)}
		{1+ \varepsilon \, N_{\rm i} \, (\tau-1)} = \hat{L}_{\rm norm}^\text{bo}(t) \, .
\end{equation}
Furthermore, by using the normalised turn variable
\begin{equation}\label{normturn}
	\bar{\tau}=\varepsilon \, N_{\rm i} \, (\tau-1) \, ,
\end{equation}
$L_{\rm norm}^\text{bo}$ can be recast in the following form 
\begin{equation}
L_{\rm norm}^\text{bo}(\bar{\tau})=
	\frac{\bar{\tau}}{1+\bar{\tau}} \, .
\label{absv1_1}
\end{equation}
Hence, $L_{\rm norm}^\text{bo}(\bar{\tau})$ has a very simple scaling law in terms of $\bar{\tau}$. This allows comparing experimental data from physics runs with different beam parameters, such as $\beta^*$, crossing angle, bunch intensity, and number of bunches (see~\cite{lumi_Part_II}). 

This treatment can be generalised to the case when the initial intensities are not the same, i.e. using the notation of the previous section, when $\xi \neq 0$. In this case, the final result reads
\begin{equation}
L_{\rm norm}^\text{bo}(\bar{\tau}) =
\frac{L_{\rm int}^\text{bo}(\bar{\tau})}{L_{\rm int}^\text{bo}(\bar{\tau}\to\infty)} = 
	\displaystyle{ \frac{1- \exp{(-\bar{\tau})}}{1-N_{\rm r}\, 
	\exp{(-\bar{\tau})}}} \,
\label{absv2}
\end{equation}
with 
\begin{equation}
L_{\rm int}^\text{bo}(\tau\to\infty) =  \frac{\Xi \, N_{\rm i,min}}{\varepsilon} \qquad 
\bar{\tau}=\varepsilon \, \xi \, \left ( \tau -1 \right ). 
\label{def}
\end{equation}
In this case the scaling law then depends on two parameters, i.e. a re-scaled turn number $\bar{\tau}$ and $N_{\rm r}$.

As mentioned in the previous section, it is easy to extend this derivation to the case where $\varepsilon=\varepsilon(\tau)$ is time-dependent by substituting $\left(\tau -1\right)\,\varepsilon \to \mu(\tau)$ as explained in \ref{app:time-dip}.

To include pseudo-diffusive effects it is enough to apply the computations made before to the general solution of the intensity-evolution equation, based on the sum of components $N^{\rm bo}_{1,2} (\tau)$ and $N^{\rm pd}_{1,2} (\tau)$, hence giving
\begin{align}
L_{\rm norm}(\tau) &= \notag
	\frac{1}{L_{\rm int}(\tau\to\infty)} \;
	\Int{1}{\tau}{\tilde{\tau}}\;\Xi
	\left[ N^{\rm bo}_{1} (\tilde{\tau}) + N^{\rm pd}_{1} (\tilde{\tau})\right]
	\left[ N^{\rm bo}_{2} (\tilde{\tau}) + N^{\rm pd}_{2} (\tilde{\tau})\right]
\, , \\
&=\label{scaling1} 
	L_{\rm norm}^{\rm bo}(\tau) +  L^\text{pd}(\tau)
\end{align}
where $L_{\rm norm}^{\rm bo}$ stands for the burn off component of the luminosity evolution derived in Eqs.~\eqref{lumint1} or~\eqref{absv2}. It is worth recalling that the terms $N^{\rm bo}_{1,2}$ and $N^{\rm pd}_{1,2}$ depend on the small parameter $\varepsilon$. It is therefore possible to find an expansion of $L^\text{pd}$ in terms of $\varepsilon$ and to truncate it at an appropriate order, which, in our case, will be the first one:
\begin{equation}\label{Lpdfitmodel}
L^{\rm pd}(\tau)=
  - N_\text{i}\,\varepsilon\Int{1}{\tau}{\tilde{\tau}}
    \left[\Exp{-\frac{D^2(\tilde{\tau})}{2}} - \Exp{-\frac{D^2(1)}{2}}\right]
    \left\{
      2 - \left[ \Exp{-\frac{D^2(\tilde{\tau})}{2}} - \Exp{-\frac{D^2(1)}{2}}\right] 
    \right \}
   + \mathcal{O}(\varepsilon^2)
\, .
\end{equation}

The generalisation to different intensities is straightforward:
\begin{multline}
L^{\rm pd}(\tau) =
	- N_\text{i,max}\;\varepsilon\Int{1}{\tau}{\tilde{\tau}}
	\left[
		\left( \Exp{-\frac{D_1^2(\tilde{\tau})}{2}} - \Exp{-\frac{D_1^2(1)}{2}} -1 \right)
\right.\times\\\left.
		\left( \Exp{-\frac{D_2^2(\tilde{\tau})}{2}} - \Exp{-\frac{D_2^2(1)}{2}} -1 \right)
		- 1
	\right] + \mathcal{O}(\varepsilon^2)
\, .
\end{multline}
In case $\varepsilon$ is time-dependent, it should be included in the integrand.
\section{Optimal physics fill duration}\label{sec:optfill}
A relevant quantity for evaluating the performance of a circular collider is the optimal fill length. A detailed discussion about the possible optimisation of this quantity has been presented in Ref.~\cite{FZ}, but there, non-linear effects were not taken into account. Indeed, the models developed in the previous section allow providing estimates of this quantity including also the impact of non-linearities. In the following, the relation 
\begin{equation}
n \, (\tau_{\rm ta} + \tau ) = \mathcal{T} 
\end{equation}
is considered, where $\tau_{\rm ta}$ is the turnaround time, i.e. the time between the end of a physics fill and the beginning of the next one, $\tau$ is the fill length that should be optimised, and $\mathcal{T}$ is the total time for physics over one year. According to the convention set in previous sections the time is always expressed in terms of turn number. With these assumptions $n$ represents the number of physics fills in one year of operation and the total integrated luminosity is given by
\begin{equation}
L_{\rm tot, norm}^{\rm bo}(\tau) = \frac{\mathcal{T}}{\tau_{\rm ta}+\tau} 
\, L_{\rm norm}^{\rm bo} (\tau) \, ,
\label{lumitotnorm}
\end{equation}
where only the burn off has been taken into account. The optimal fill length $\tau_{\rm fill}$ can be obtained by setting to zero the derivative of $L_{\rm tot, norm}^{\rm bo}$ with respect to $\tau$ and it reads
\begin{equation}
\tau_{\rm fill}^{\rm bo} 
\begin{cases} =
\displaystyle{1+\sqrt{\frac{1+\tau_{\rm ta}}{\varepsilon \, N_{\rm i}}}} \approx \displaystyle{\sqrt{\frac{\tau_{\rm ta}}{\varepsilon \, N_{\rm i}}}}& 
\qquad \xi = 0  \\
& \\ \label{optfillscale} 
\approx 
\displaystyle{1+\sqrt{\frac{1+\tau_{\rm ta}}{\varepsilon \, N_{\rm i, min}}}}
\approx \displaystyle{\sqrt{\frac{\tau_{\rm ta}}{\varepsilon \, N_{\rm i, min}}}}& 
\qquad \xi \neq 0
\end{cases}
\end{equation}
where the exponential terms in Eq.~(\ref{absv2}) have been approximated considering that the arguments are small and the power series has been truncated to first order in $\varepsilon$. Clearly, the expression for $\xi \neq0$ in Eq.~(\ref{optfillscale}) tends to that for $\xi =0$.
\begin{figure}[htb]
  \begin{center}
      \includegraphics[width=0.55\linewidth,clip=]{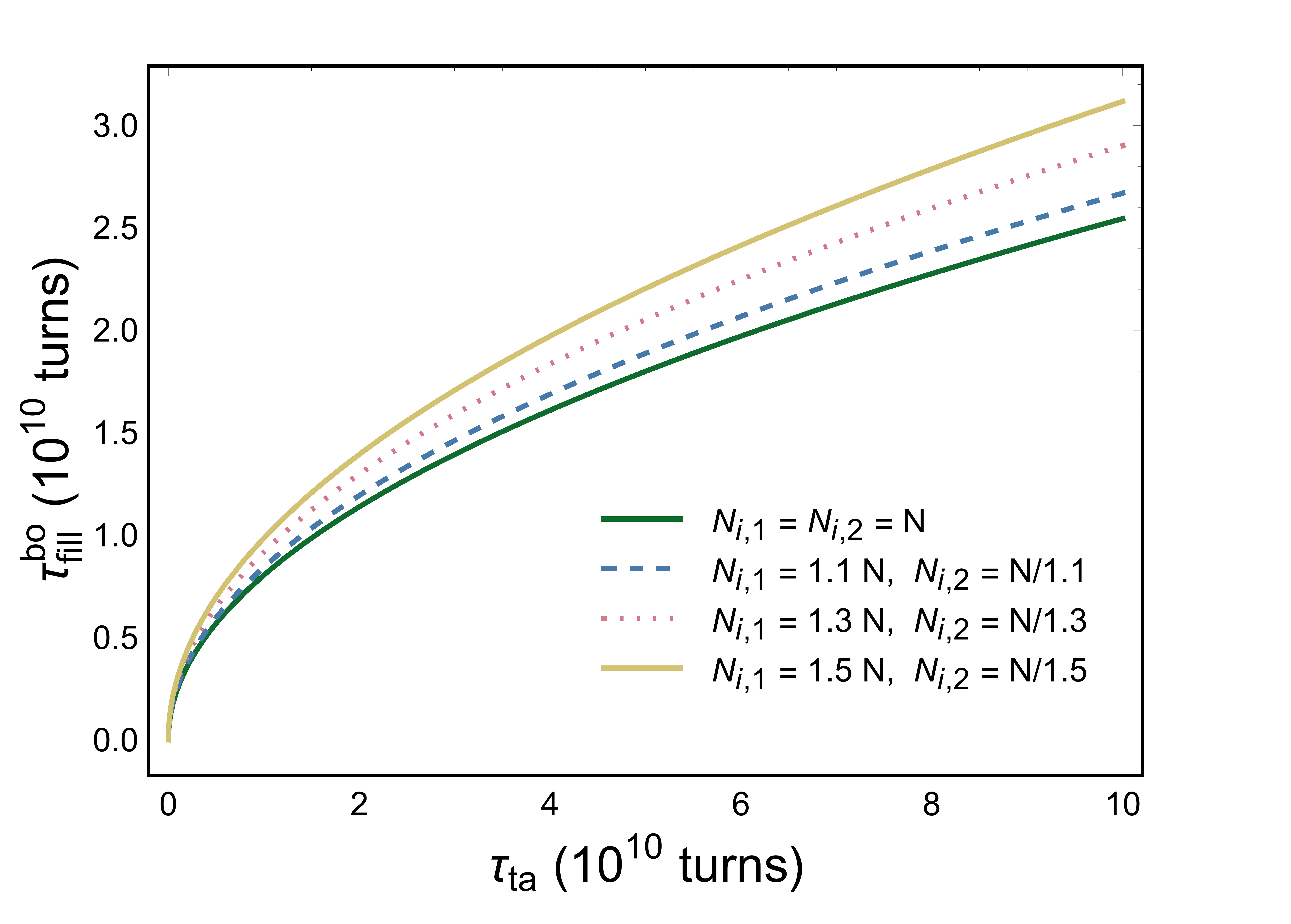}
  \end{center}
  \caption{Dependence of $\tau_{\rm fill}^{\rm bo}$ on $\tau_{\rm ta}$ for equal and unequal intensities.}
  \label{example_fill}
\end{figure}
Equations~(\ref{optfillscale})  provide the scaling laws of the optimal fill length when the \mbox{burn off} is the mechanism for intensity reduction. Indeed, $\tau_{\rm fill}^{\rm bo}$ scales with intensity and the turnaround time, but only as the square root. As an example, in Fig.~\ref{example_fill} the optimal fill length is shown as a function of $\tau_{\rm ta}$ for different configurations with equal and unequal beam intensities. Also in this case $N_{\rm i, 1} N_{\rm i, 2} = N^2$ has been set, to allow for a direct comparison of the various configurations of the constraint.

It is also possible to estimate how $L_{\rm tot, norm}^{\rm bo}(\tau_{\rm fill}^{\rm bo})$ scales with respect to, e.g. $\tau_{\rm ta}$, thus obtaining 
\begin{equation}
L_{\rm tot, norm}^{\rm bo}(\tau_{\rm fill}^{\rm bo})  \propto 
\begin{cases}
\mathcal{T} \varepsilon \, N_{\rm i} \left [ 1 - 2 \, 
\sqrt{\varepsilon \, N_{\rm i} \, \tau_{\rm ta}} \right ] & 
\text{if $\tau_{\rm ta} \to 0$}, \\
& \\
\displaystyle{\frac{\mathcal{T}}{\tau_{\rm ta}}} & 
\text{if $\tau_{\rm ta} \to \infty$}.
\end{cases}
\end{equation}
The increase in $L_{\rm tot, norm}^{\rm bo}$ generated by a reduction of $\tau_{\rm ta}$ scales only as the square root of $\tau_{\rm ta}$, whereas $L_{\rm tot, norm}^{\rm bo}$ decreases with increasing $\tau_{\rm ta}$ as the inverse of the turnaround time. All this indicates how easily the performance of a collider can be spoilt. The value of $L_{\rm tot, norm}^{\rm bo}(\tau_{\rm fill}^{\rm bo})$ depends on $N_{\rm i}$, which represents either the common beam intensity or the highest one. 

The next step is to evaluate how the term $L^{\rm pd}$ changes the conclusions concerning the optimal fill time. In this case the total normalised luminosity reads
\begin{equation}
\begin{split}
L_{\rm tot, norm}(\tau) & =
  \frac{\mathcal{T}}{\tau_{\rm ta}+\tau} \,
  \left[ L_{\rm norm}^{\rm bo} (\tau) + L^{\rm pd}(\tau) \right] \\
& =
  \underbrace{
    \frac{\mathcal{T}}{\tau_{\rm ta}+\tau} \, L_{\rm norm}^{\rm bo} (\tau)
  }_{
    \displaystyle{L_{\rm tot, norm}^{\rm bo}(\tau)}
  }
  +\underbrace{
    \frac{\mathcal{T}}{\tau_{\rm ta}+\tau}\, L^{\rm pd}(\tau)
  }_{
    \displaystyle{L_{\rm tot, norm}^{\rm pd}(\tau)}
  } \\
\end{split}
\end{equation}
where $L_{\rm tot, norm}^{\rm bo}(\tau)$ is the term for which the optimal fill time is $\tau_{\rm fill}^{\rm bo}$. 

The value $\tau_{\rm fill}$ that maximises $L_{\rm tot, norm}(\tau)$ is obtained by setting to zero the derivative of $L_{\rm tot, norm}(\tau)$. An approximate expression can be obtained by developing the first-order derivatives of $L_{\rm tot, norm}^{\rm bo}(\tau)$ and $L_{\rm tot, norm}^{\rm pd}(\tau)$ taking into account that, by definition, $\dot{L}_{\rm tot, norm}^{\rm bo}(\tau_{\rm fill}^{\rm bo})=0$. The expression, then, reads
\begin{equation}
\tau_{\rm fill} \approx
  \tau_{\rm fill}^{\rm bo} -
  \frac{
    \dot{L}_{\rm tot, norm}^{\rm pd}(\tau_{\rm fill}^{\rm bo})
  }%
  {
    \ddot{L}_{\rm tot, norm}^{\rm bo}(\tau_{\rm fill}^{\rm bo})
    +\ddot{L}_{\rm tot, norm}^{\rm pd}(\tau_{\rm fill}^{\rm bo})
  } \, .
\label{taufill}
\end{equation}
Equation~\eqref{taufill} can be used to derive approximate expressions for $\Delta \tau_{\rm fill} = \tau_{\rm fill} - \tau_{\rm fill}^{\rm bo}$ and the relative error, namely $\Delta \tau_{\rm fill}/\tau_{\rm fill}^{\rm bo}$, which reads
\begin{equation}
\frac{\Delta \tau_{\rm fill}}{\tau_{\rm fill}^{\rm bo}} =
  -\frac{
    \dot{L}_{\rm tot, norm}^{\rm pd}(\tau_{\rm fill}^{\rm bo})
  }%
  {
    \tau_{\rm fill}^{\rm bo} \,
    \left[
      \ddot{L}_{\rm tot, norm}^{\rm bo}(\tau_{\rm fill}^{\rm bo})
      +\ddot{L}_{\rm tot, norm}^{\rm pd}(\tau_{\rm fill}^{\rm bo})
    \right]
  } \, .
\end{equation}
\section{Conclusions\label{sec:conclusions}}
In this paper, a model for luminosity evolution based on the scaling law of dynamic aperture as a function of time, in addition to particle burn off, has been presented and discussed in detail. 

Estimates of the luminosity evolution have been worked out, together with prediction for the scaling law of the integrated luminosity as a function of the fill length and of the optimal fill length. While most physical parameters have been considered as being time-independent, generalisation to cases in which these parameters vary with time is provided in the appendices. Non-negligible differences with respect to simpler models based on burn off only have been clearly found. 

Application of the proposed approach to collider's data is discussed in a companion paper~\cite{lumi_Part_II}, where a subset of the LHC Run~1 data is analysed and discussed in detail.
\section*{Acknowledgements}
One of the authors (MG) would like to thank A.~Bazzani for interesting discussions. Moreover, he would also like to acknowledge the interesting and stimulating remarks and comments from G.~Arduini, V.~Lebedev, S.~Nagaitsev, V.~Shiltsev, L.~Vorobief, T.~Zolkin.
\section*{References}
\clearpage
\appendix
\section{Refinement of the burn off model\label{app:refinement}}
The model that includes only the burn off can be improved to take into account also intensity-independent emittance variation, i.e. either growth or damping. These phenomena include beam-residual gas scattering and radiation damping. Here, the following form for the variation of the normalised emittance~\footnote{It is worth stressing that throughout the paper the concept of emittance always refers to the normalised emittance. This has an effect on the values of the standard expressions for the coefficients $\lambda_i$.} is considered:
\begin{equation}
\dot{\epsilon^*}=-\lambda_1 \, \epsilon^* + \lambda_2 , \qquad \text{where} 
\qquad \lambda_i > 0,
\label{emit_evol1}
\end{equation}
where the term $-\lambda_1 \epsilon^* $ describes a damping of the beam emittance, and $\lambda_2$ a growth. The emittance evolution is given by the following expression
\begin{equation}
\epsilon^*(\tau)=\frac{\lambda_2+\Exp{\lambda_1 \,( 1- \tau)} %
\, \left(\lambda_1 \, \epsilon^*_1-\lambda_2\right)}{\lambda_1}
\label{emevol}
\end{equation}
where $\epsilon^*_1=\epsilon^*(1)$. The combined effect of damping and growth makes it possible to define an equilibrium emittance $\epsilon^*_{\rm eq}$ given by
\begin{equation}
\epsilon^* _{\rm eq} =\frac{\lambda_2}{\lambda_1}
\end{equation}
so that Eq.~(\ref{emevol}) can be recast in the form
\begin{equation}
\epsilon^*(\tau)=\epsilon^*_{\rm eq} + \Exp{\lambda_1 \,( 1- \tau)} \left ( 
\epsilon^*_1-\epsilon^*_{\rm eq} \right )
\label{emit_evol2}
\end{equation}
and $1/\lambda_1$ gives the time over which the equilibrium value of the emittance is reached.

The relative variation of the normalised emittance is readily computed to be 
\begin{equation}
\frac{\epsilon^*(\tau)-\epsilon^*_1}{\epsilon^*_1} = 
\left [ \Exp{\lambda_1 \,( 1- \tau)}-1 \right ] \, 
\left ( 1 - \frac{\epsilon^*_{\rm eq}}{\epsilon^*_1 }\right ).
\end{equation}

Under the assumptions mentioned at the beginning of this section, it is possible to give the expressions of $\lambda_i$ as~\cite{handbook,bryant}
\begin{eqnarray}
\lambda_1 & = & 
\frac{C_{\gamma} \, \left ( m_{\rm p} c^2 \right )^3}{2 \, \pi} 
\frac{\gamma_{\rm r}^2}{\beta_{\rm r}} \, \mathcal{I}_2 \qquad 
\mathrm{with} \qquad 
C_{\gamma} =  \frac{4 \, \pi}{3} \frac{r_{\rm p}}%
{\left ( m_{\rm p} c^2\right )^3}\, , \qquad 
\mathcal{I}_2 = \Oint{}{}{s}\frac{1}{\rho_x^2} \nonumber \\
& & \\
\lambda_2 & = & \lambda_{2,\rm rad} + \lambda_{2, \rm scat} \nonumber
\end{eqnarray}
where $r_{\rm p}, m_{\rm p} c^2, E, \rho_x$ are the classical proton radius, the proton mass, energy, and bending radius, respectively. Typical values for the LHC at $4$~TeV are $\lambda_1 \approx 7.52 \times 10^{-10}$. Furthermore, the components of $\lambda_2$ can be expressed as~\cite{handbook,bryant}
\begin{equation}
\lambda_{2,\rm rad} = \frac{55}{48 \, \sqrt 3} \, 
\frac{\hbar c}{m_{\rm p} c^2 }  \, 
\frac{\gamma_{\rm r}^4} {\beta_{\rm r}\, f_{\rm rev}} \, \mathcal{I}_5
\end{equation}
with
\begin{equation}
\mathcal{I}_5 = \Oint{}{}{s}\frac{\mathcal{H}}{|\rho_x|^3}\qquad \qquad
\mathcal{H}(s) = \beta_x {D'_x}^2 + 2\, \alpha_x D_x \, {D'}_x + 
\gamma_x {D_x}^2 \, , 
\end{equation}
where $\beta_x, \alpha_x, \gamma_x$ are the Twiss parameters~\cite{Courant-Snyder} and $D_x, {D'}_x$ the horizontal dispersion and its derivative, respectively. For typical LHC parameters corresponding to the runs at lower energies in 2011 and 2012, $\lambda_{2,\rm rad} \approx 1.38 \times 10^{-23}$~m, hence completely negligible.

The second component of $\lambda_2$ is expressed by 
\begin{equation}
\lambda_{2, \rm scat} = 4 \, \pi^2 r_{\rm p}^2 \, c \, 
\left ( \frac{m_{\rm p} c}{p} \right )^2 
\frac{1}{\beta_{\rm r} \, \gamma_{\rm r} \, f_{\rm rev}} \, \frac{N_a}{R \, T} \, P 
\, \langle \beta \rangle \, 
Z^2 \ln \left [ \frac{3.84 \times 10^4}{\left ( A \, Z \right )^{1/3}}\right ]
\end{equation}
where $N_a, R, T, P$ stand for the Avogadro number, the gas constant, the temperature in Kelvin, and the pressure in Pascal, respectively. The properties of the residual gas are taken into account via $A$ and $Z$, the atomic and the charge number, respectively. Finally, the impact of the ring's optics is taken into account by the term $\langle \beta \rangle$ representing the average beta-function, while $p$ is beam momentum. 

\begin{table}[htb]
\centering
\caption{Summary of the pressure at $5$~K and other parameters relevant for the computation of $\lambda_{2,\rm scat}$ for the various gas species in the LHC for nominal conditions from Ref.~\cite{LHCDR}.}
\begin{tabular}{lccc}
\hline
Gas & Pressure        & $Z^2 \, \displaystyle{\ln \left [ %
\frac{3.84 \times 10^4}{\left ( A \, Z \right )^{1/3}}\right ]}$ & $\lambda_{2,\rm scat}$ \\
    & ($ 10^{-9}$ Pa) &                                    & ($ 10^{-23}$ m)\\
\hline
H$_2$  &  $ 67.0  $ & $\phantom{11}21.1$ & $\phantom{1}4.3 $ \\
He     &  $ 51.0  $ & $\phantom{11}39.4$ & $\phantom{1}6.2 $ \\
CH$_4$ &  $ 11.0  $ & $\phantom{1}370.9$ & $12.5 $ \\
H$_2$O &  $ 11.0  $ & $\phantom{1}575.2$ & $19.4 $ \\
CO     &  $\phantom{1}7.5 $ & $\phantom{1}884.5$ & $20.2 $ \\
CO$_2$ &  $\phantom{1}4.9 $ & $1436.8$           & $21.5 $ \\
\hline
\label{scatt}
\end{tabular}
\end{table}

In Table~\ref{scatt} the parameters corresponding to the various gas species in the LHC for nominal conditions~\cite{LHCDR} are listed, together with the corresponding values of $\lambda_{2,\rm scat}$.

It is now possible to verify that the condition $\epsilon^*_{\rm eq} / \, \epsilon^*_1 <<1$, required to neglect the contribution of $\lambda_2$ with respect to that of $\lambda_1$ in the emittance evolution, is always satisfied at the LHC with nominal parameters from Table~\ref{scatt}. Moreover, it is also easy to check that even the impact of $\lambda_1$, assuming Run~1 beam parameters, can be correctly neglected. Nonetheless, for the sake of completeness, the complete discussion of the considered emittance growth model is reported in \ref{app:time-dip}.
\section{Evolution of beam intensity and luminosity with burn off and additional time dependency \label{app:time-dip}}
Let us consider the homogeneous part of Eq.~(\ref{int_main2}) assuming that the luminosity is changing not only due to particles' burn off, but also for other time-dependent effects, such as emittance growth. In this case it is possible to recast the equation in the form
\begin{equation}
\begin{cases}
\dot{N}_1(\tau) & = -\varepsilon(\tau) \, N_1(\tau) \, N_2(\tau) \\
\dot{N}_2(\tau) & = -\varepsilon(\tau) \, N_1(\tau) \, N_2(\tau). 
\end{cases}
\label{int_main3}
\end{equation}
Once more, the solutions are of two types, as they can satisfy
\begin{equation}
\begin{cases}
N_1(\tau)  & = \phantom{-} N_2(\tau) \\
\dot{N}_2(\tau) & = - \varepsilon(\tau) \, N_2^2(\tau) \, ,
\end{cases}
\label{ssol3_1}
\end{equation}
or
\begin{equation}
\begin{cases}
  N_1(\tau)  & = \phantom{-} N_2(\tau) + \xi \\
\dot{N}_2(\tau) & = -\varepsilon(\tau) \, 
\left [ N_2(\tau) + \xi \right ] \, N_2(\tau) \, .
\end{cases}
\label{ssol3_2}
\end{equation}
Equation~(\ref{ssol3_1}) can be solved using standard methods, giving
\begin{equation}
\begin{cases}
N_1(\tau)  & = \phantom{-} N_2(\tau) \\[0.5em]
N_2(\tau) & = \displaystyle \frac{N_{\rm i}}{1+N_{\rm i} \, \mu(\tau)}\, ,
\end{cases}
\label{ssol3_1_1}
\end{equation}
where $N_1(1)=N_2(1)=N_{\rm i}$, and
\begin{equation}\label{eq:mu}
  \mu(\tau) =
    \Int{1}{\tau}{\tilde{\tau}} \varepsilon(\tilde{\tau}) \,.
\end{equation}
Note that this solution has the same structure as the time-independent case in Eq.~\eqref{ssol1} after making the substitution
\begin{equation}
  (\tau-1)\,\varepsilon \to \mu(\tau)\,.
\end{equation}

Equation~(\ref{ssol3_2}) can also be solved with standard methods and one obtains
\begin{equation}
\begin{cases}
N_1(\tau) & =
	\xi \displaystyle{\frac{1}{1 - N_{\rm r}\,\Exp{-\xi\,\mu(\tau)}}} \\[2.5em]
N_2(\tau) & =
	\xi \displaystyle{\frac{N_{\rm r}\,\Exp{-\xi\,\mu(\tau)}}
		{1 - N_{\rm r}\,
			\Exp{-\xi \,\mu(\tau)}
		}
	}
\end{cases}
\label{ssol3_2_1}
\end{equation}
which again holds for $\xi \neq 0$ independently of the sign of $\xi$. 

The possibility of expressing $N_j(\tau)$ in closed form depends then on the possibility of evaluating $ \Int{}{}{\tau} \varepsilon(\tilde{\tau})$ analytically.

The computation of the instantaneous luminosity does not pose any problem, apart from the involved expression. On the other hand, the expression of the integrated luminosity deserves some comments. In the case under consideration one has
\begin{equation}
L_{\rm int} (\tau) = \Int{1}{\tau}{\tilde{\tau}} L(\tilde{\tau}) = 
  \frac{f_\text{rev}}{\sigma_{\rm int} \, n_{\rm c}} \,
  \Int{1}{\tau}{\tilde{\tau}}
  N_1(\tilde{\tau}) \, N_2(\tilde{\tau}) \, \varepsilon(\tilde{\tau})
\label{int_lumi_app}
\end{equation}
and it is immediate to notice by inspecting Eqs.~(\ref{ssol3_1_1}) and (\ref{ssol3_2_1}) that the dependence of $N_1, N_2$ on $\tau$ is only via the term $\mu(\tau)$. Therefore, the change of variable $\mu(\tau)=\Intnl{1}{\tau}{\tilde{\tau}} \varepsilon(\tilde{\tau})$ in the integral allows simplifying Eq.~(\ref{int_lumi_app}) thus obtaining for the burn off contribution
\begin{equation}
L_{\rm int}^\text{bo} (\tau) =
\begin{cases}
 \displaystyle{\frac{N_{\rm i}\,f_\text{rev}}{\sigma_{\rm int} \, n_{\rm c}}} \, 
\frac{N_{\rm i} \, \mu(\tau)} {1+N_{\rm i} \, \mu(\tau)}  
& 
	\xi = 0
\\[2.5em]
\displaystyle{
	\frac{N_{\rm i,min}\,f_\text{rev}}{\sigma_{\rm int} \, n_{\rm c}}} \, 
	\frac{1-\Exp{-\xi\, \mu(\tau)}}%
	{1 - N_{\rm r} \, \Exp{-\xi\, \mu(\tau)}}
&
	\xi \neq 0
\, ,
\end{cases}
\label{int_lumi_app1}
\end{equation}
where $N_{\rm r}$ has been defined before, i.e. $N_{\rm r}=N_{\rm i,min}/N_{\rm i,max}$. Once more, the possibility of evaluating in closed form the integral of $\varepsilon(\tau)$ is the condition to find a closed form expression for $\mu(\tau)$ and hence for the integrated luminosity. Normalisation factors for the integrated luminosity can be found as
\begin{equation}
\begin{cases}
 \displaystyle{\frac{N_{\rm i}\,f_\text{rev}}{\sigma_{\rm int} \, n_{\rm c}}} 
&  \qquad \xi = 0  \\
& \\
 \displaystyle{\frac{N_{\rm i,min}\,f_\text{rev}}{\sigma_{\rm int} \, n_{\rm c}}}
& \qquad \xi \neq 0  \, ,
 \end{cases}
\label{int_lumi_norm}
\end{equation}
so that a very simple expression for $L_{\rm norm}^\text{bo} (\tau)$ can be found, namely
\begin{equation}
L_{\rm norm}^\text{bo} (\tau) =
\begin{cases}
\displaystyle
\frac{N_{\rm i} \, 
\mu(\tau)}%
{1+N_{\rm i} \, 
\mu(\tau)}  
& 
	\xi = 0
\\[2.5em]
\displaystyle
\frac{1 - \Exp{-\xi\,\mu(\tau)}}%
	{1 - N_{\rm r} \, \Exp{-\xi\,\mu(\tau)}}
&
	\xi \neq 0
\, .
\end{cases}
\label{int_lumi_app2}
\end{equation}
It is readily seen that the expressions for the normalised integrated luminosity feature rather simple scaling laws with respect to $ \mu(\tau)$, very much like the time-independent case.

The possibility of finding an expression for the optimal physics fill length can also be explored. To this aim the total luminosity is given by
\begin{equation}
L_{\rm tot, norm}(\tau)= \frac{\mathcal{T}}{\tau_{\rm ta}+\tau} \, L_{\rm norm}(\tau),
\end{equation}
where the symbols have the same meaning as in the main body of the paper. The optimal fill length can be computed by looking for the solutions of the equation:
\begin{equation}
\begin{split}
\frac{\dif L_{\rm tot, norm}(\tau)}{\dif \tau} & = -\frac{\mathcal{T}}%
{\left ( \tau_{\rm ta} + \tau\right )^2} \, L_{\rm norm} (\tau) + 
\frac{\mathcal{T}}{\tau_{\rm ta} + \tau} \, 
\frac{\dif L_{\rm norm}(\mu(\tau))}{\dif \mu} \, 
\frac{\dif \mu (\tau)}{\dif \tau} \\
& =  -\frac{\mathcal{T}}%
{\left ( \tau_{\rm ta} + \tau\right )} \left [ \frac{1}%
{\left ( \tau_{\rm ta} + \tau\right )} \, L_{\rm norm} (\tau) - 
\frac{\dif L_{\rm norm}(\mu(\tau))}{\dif \mu} \, \varepsilon(\tau) \right ]= 0 \, .
\end{split}
\end{equation}
The derivative of the burn off contribution to the integrated luminosity can be computed to be 
\begin{equation}
\frac{\dif L_{\rm norm}^{\text{bo}} (\mu(\tau))}{\dif \mu} =
\begin{cases}
 \displaystyle{\frac{N_{\rm i}}%
{\left (1+N_{\rm i} \, 
\displaystyle{\mu(\tau)} \right )^2}}
& \xi = 0  \\[2.5em]
\displaystyle
\frac{\xi^2}{N_{\rm i,max}} \,
\frac{\Exp{-\xi \,\mu(\tau) }}%
  {\left( 1 - N_{\rm r} \,  \Exp{-\xi \, \mu(\tau)} \right)^2}
& \xi \neq 0  \, .
 \end{cases}
\label{int_lumi_der}
\end{equation}
Therefore, the optimal fill time $\tau_{\rm fill}^\text{bo}$ (taking only the burn off into account) is the solution of the following equations
\begin{equation}
\begin{cases}
\left ( \tau_{\rm ta} + \tau_{\rm fill}^\text{bo} \right ) \, \varepsilon(\tau_{\rm fill}^\text{bo})- 
\Big( 1 + N_{\rm i} \, \mu(\tau_{\rm fill}^\text{bo}) \Big) \, \mu(\tau_{\rm fill}^\text{bo})
  & = 0  \qquad \xi = 0  \\[1.5em]
\left ( \tau_{\rm ta} + \tau_{\rm fill}^\text{bo} \right ) \,\xi^2\, \varepsilon(\tau_{\rm fill}^\text{bo})
+ N_{\rm i,max} \left( 1 - \Exp{\xi\, \mu(\tau_{\rm fill}^\text{bo})} \right) + \\
\phantom{\left ( \tau_{\rm ta} + \tau_{\rm fill}^\text{bo} \right ) \,\xi^2\, \varepsilon(\tau_{\rm fill}^\text{bo})}
+ N_{\rm i,min} \left( 1 - \Exp{-\xi\,\mu(\tau_{\rm fill}^\text{bo})} \right)
& = 0 \qquad \xi \neq 0  \, .
 \end{cases}
\label{opt_fill}
\end{equation}

The model introduced in Eqs.~(\ref{emit_evol1}) and~(\ref{emit_evol2}) can be included in the estimate of the intensity evolution and, eventually, in the expression for the luminosity and its integral. If the change in the geometrical factor $F$ over the lapse of time of a physical fill is negligible, then a closed form for $\mu(\tau)$ can be found. By defining
\begin{equation}
\varepsilon (\tau) = \frac{\varepsilon_1}{\epsilon^*(\tau)} \qquad
\varepsilon_1 = \frac{\sigma_{\rm int} \, n_{\rm c}}{f_{\rm rev}} 
\left ( \frac{L}{N_{\rm i,1} \, N_{\rm i, 2}} \right ) \, \epsilon^*_1
\end{equation}
then 
\begin{equation}
\mu(\tau) =
  \varepsilon_1 \, \Int{1}{\tau}{\tilde{\tau}} \frac{1}{\epsilon^*(\tilde{\tau})}
=
  \frac{\varepsilon_1}{\lambda_1 \, \epsilon^*_{\rm eq}} \, 
 \log {\left \{ 1 + \frac{\epsilon^*_{\rm eq}}{\epsilon^*_1} \, 
\left [ \Exp{\lambda_1 (\tau -1)}-1 \right] \right \}} \, .
\end{equation}
According to Eqs.~(\ref{ssol3_1_1}) and~(\ref{ssol3_2_1}) the intensity evolution reads
\begin{equation}
\begin{cases}
N_1^\text{bo}(\tau)  & = \phantom{-} N_2^\text{bo}(\tau) \\
N_2^\text{bo}(\tau) & = \displaystyle{\frac{N_{\rm i}}%
{1+ \displaystyle{\frac{N_{\rm i} \, \varepsilon_1}%
{\lambda_1 \, \epsilon^*_{\rm eq}} \, 
\log \left \{1 - \frac{\epsilon^*_{\rm eq}}{\epsilon^*_1} 
\left [ 1 - \Exp{\lambda_1 (\tau -1)} \right ] % 
\right \}}}},
\end{cases}
\end{equation}
for the case of equal intensity and
\begin{equation}
\begin{cases}
N_1^\text{bo}(\tau)  & = \displaystyle{%
\xi
\frac{1}{1-N_{\rm r} \, 
	\displaystyle{\left \{1 - \frac{\epsilon^*_{\rm eq}}{\epsilon^*_1} %
\left [ 1 - \Exp{\lambda_1 (\tau -1)} \right ] \right \}^{-\xi\,\varepsilon_1 /
(\lambda_1 \, \epsilon^*_{\rm eq}) }}}} \\[3.5em]
N_2^\text{bo}(\tau)  & = \displaystyle{%
\xi
\frac{N_{\rm r} \, 
	\displaystyle{\left \{1 - \frac{\epsilon^*_{\rm eq}}{\epsilon^*_1} %
\left [ 1 - \Exp{\lambda_1 (\tau -1)} \right ] \right \}^{-\xi\,\varepsilon_1 /
(\lambda_1 \, \epsilon^*_{\rm eq}) }}}
{1-N_{\rm r} \, 
	\displaystyle{\left \{1 - \frac{\epsilon^*_{\rm eq}}{\epsilon^*_1} %
\left [ 1 - \Exp{\lambda_1 (\tau -1)} \right ] \right \}^{-\xi\,\varepsilon_1 /
(\lambda_1 \, \epsilon^*_{\rm eq}) }}}}
\, ,
\end{cases}
\end{equation}
whenever the initial intensities are different. It is worth noting that the time-dependence of the beam intensity becomes much more involved when emittance evolution is included in the model. 

The normalisation derived in this Appendix can be applied and by using Eq.~(\ref{int_lumi_app2}) one obtains
\begin{equation}
L_{\rm norm}^\text{bo} (\tau) =
\begin{cases}
\frac{\displaystyle{\frac{N_{\rm i} \, 
\varepsilon_1}{\lambda_1 \, \epsilon^*_{\rm eq}} \, 
 \log {\left \{ 1+ \frac{\epsilon^*_{\rm eq}}{\epsilon^*_1} \, 
\left [ \Exp{\lambda_1 (\tau -1)}-1 \right ] \right \}}}}%
{1+ 
\displaystyle{\frac{N_{\rm i} \, \varepsilon_1}{\lambda_1 \, 
\epsilon^*_{\rm eq}} \, 
 \log {\left \{ 1+ \frac{\epsilon^*_{\rm eq}}{\epsilon^*_1} \, 
\left [ \Exp{\lambda_1 (\tau -1)}-1 \right ] \right \}}}}  
& \xi = 0  \\
& \\
\displaystyle{\frac{1-\displaystyle{\left \{ 1+
\frac{\epsilon^*_{\rm eq}}{\epsilon^*_1} \, \left [ \Exp{\lambda_1 (\tau -1)}-1 \right ]
 \right \}}^{-\eta_1}}
{1 - N_{\rm r} \, \displaystyle{\left \{ 1+ 
\frac{\epsilon^*_{\rm eq}}{\epsilon^*_1} \, \left [ \Exp{\lambda_1 (\tau -1)}-1 \right ]
 \right \}}^{-\eta_1}}} \qquad 
\eta_1 = \xi\frac{\varepsilon_1}{\lambda_1 \, \epsilon^*_{\rm eq}} & \xi > 0
\, .
 \end{cases}
\label{int_lumi_td}
\end{equation}

The simple expression~\eqref{absv1_1} can be retrieved by using a change of variable of the type
\begin{equation}
\bar{\tau}= \frac{N_{\rm i} \, \varepsilon_1}{\lambda_1 \, 
\epsilon^*_{\rm eq}} \, \log {\left \{ 1+ \frac{\epsilon^*_{\rm eq}}%
{\epsilon^*_1} \, \left [ \Exp{\lambda_1 (\tau -1)}-1 \right ] \right \}} \, ,
\label{int_lumi_td1}
\end{equation}
which should replace the choice for the new variable $\bar{\tau}$ made in Eq.~\eqref{normturn}. It is also worth noting that whenever time-dependent effects are included, transcendental functions are appearing in the expression of $L_{\rm norm}(\tau)$.

The estimate of the optimal physics fill time $\tau_{\rm fill}$ by means of Eq.~(\ref{opt_fill}) can be done only numerically. As a last remark, it is worth stressing that the split of the luminosity evolution into the sum of two parts according to Eq.~(\ref{scaling1}) remains unchanged.
\section{Derivation of the proposed model\label{app:derivation}}
Equations~(\ref{int_main2}) can be cast in the form of a Riccati equation~\cite{riccati}, although this does not provide any useful information about the properties of its solutions. Therefore, a different approach has been applied, i.e. constructing the solution as the sum of two components, namely
\begin{equation}
N_j(\tau)=N^{\rm bo}_j(\tau)+N^{\rm pd}_j(\tau) \, ,
\end{equation} 
which, applied to Eq.~\eqref{int_main2} gives
\begin{equation}
\begin{cases}
\dot{N}^{\rm pd}_1(\tau) & = -\varepsilon \, \left [ N^{\rm bo}_1(\tau) \, N^{\rm pd}_2(\tau) + N^{\rm bo}_2(\tau) \, N^{\rm pd}_1(\tau) + N^{\rm pd}_1(\tau) \, N^{\rm pd}_2(\tau) \right ] + \\
& \phantom{=} - \mathcal{D}_1(\tau) \\
\dot{N}^{\rm pd}_2(\tau) & = -\varepsilon \, \left [ N^{\rm bo}_1(\tau) \, N^{\rm pd}_2(\tau) + N^{\rm bo}_2(\tau) \, N^{\rm pd}_1(\tau) + N^{\rm pd}_1(\tau) \, N^{\rm pd}_2(\tau) \right ] + \\
& \phantom{=} - \mathcal{D}_2(\tau) \, .
\end{cases}
\label{int_main4}
\end{equation}
By replacing the developments in power series of $N^{\rm bo}_j(\tau), N^{\rm pd}_j(\tau)$ 
\begin{equation}
 N^{\rm bo}_j(\tau) = \sum_{m=0}^{+\infty} \mathcal{N}^{\rm bo}_{m,j}(\tau) \, \varepsilon^m \, , \qquad 
 N^{\rm pd}_j(\tau)= \sum_{m=0}^{+\infty} \mathcal{N}^{\rm pd}_{m,j}(\tau) \, \varepsilon^m 
\, , \qquad j=1,2 \, .
\label{sol1}
\end{equation}
into Eq.~(\ref{int_main4}) and using the Cauchy product of series, one obtains 
\begin{equation}
\begin{cases}
\displaystyle{\sum_{m=0}^{+\infty} \varepsilon^{m} \dot{\mathcal{N}}^{\rm pd}_{m,1}(\tau)} 
& = - 
\displaystyle{\sum_{m=0}^{+\infty}} \varepsilon^{m+1} \, %\left \{ 
\displaystyle{\sum_{k=0}^m} \left [ \mathcal{N}^{\rm bo}_{k,1}(\tau) \, \mathcal{N}^{\rm pd}_{m-k,2}(\tau)+ \right . \\ 
& \phantom{=} + \left . \mathcal{N}^{\rm bo}_{k,2}(\tau) \, \mathcal{N}^{\rm pd}_{m-k,1}(\tau)+ \mathcal{N}^{\rm pd}_{k,1}(\tau) \, \mathcal{N}^{\rm pd}_{m-k,2}(\tau) \right ] -\mathcal{D}_1(\tau)%} 
\\
& \\
\displaystyle{\sum_{m=0}^{+\infty} \varepsilon^m \dot{\mathcal{N}}^{\rm pd}_{m,2}(\tau)}
& = - 
\displaystyle{\sum_{m=0}^{+\infty}} \varepsilon^{m+1} \, %\left \{ 
\displaystyle{\sum_{k=0}^m} \left [ \mathcal{N}^{\rm bo}_{k,1}(\tau) \, \mathcal{N}^{\rm pd}_{m-k,2}(\tau)+ \right . \\
& \phantom{=} + \left . \mathcal{N}^{\rm bo}_{k,2}(\tau) \, \mathcal{N}^{\rm pd}_{m-k,1}(\tau)+ \mathcal{N}^{\rm pd}_{k,1}(\tau) \, \mathcal{N}^{\rm pd}_{m-k,2}(\tau) \right ] -\mathcal{D}_2(\tau) \\
\end{cases}
%\label{ssol2}
\end{equation}
and the following recursive equations are found
\begin{equation}
\begin{cases}
  \dot{\mathcal{N}}^{\rm pd}_{0,1}(\tau) & = -\mathcal{D}_1(\tau) \\
  \dot{\mathcal{N}}^{\rm pd}_{0,2}(\tau) & = -\mathcal{D}_2(\tau)
\end{cases}
\label{sol1_1}
\end{equation}
and 
\begin{equation}
\begin{cases}
  \dot{\mathcal{N}}^{\rm pd}_{m,1}(\tau) & = -\displaystyle{\sum_{k=0}^{m-1}}
\mathcal{N}^{\rm bo}_{k,1}(\tau) \, \mathcal{N}^{\rm pd}_{m-(k+1),2}(\tau)+\mathcal{N}^{\rm bo}_{k,2}(\tau) \, \mathcal{N}^{\rm pd}_{m-(k+1),1}(\tau) + \\
& \phantom{=} + \mathcal{N}^{\rm pd}_{k,1}(\tau) \, \mathcal{N}^{\rm pd}_{m-(k+1),2}(\tau) \\
& \qquad \qquad \qquad \qquad \qquad \qquad \qquad \qquad \qquad \qquad \qquad \qquad \qquad m \geq 1 \\
  \dot{\mathcal{N}}^{\rm pd}_{m,2}(\tau) & = -\displaystyle{\sum_{k=0}^{m-1}} 
\mathcal{N}^{\rm bo}_{k,1}(\tau) \, \mathcal{N}^{\rm pd}_{m-(k+1),2}(\tau)+\mathcal{N}^{\rm bo}_{k,2}(\tau) \, \mathcal{N}^{\rm pd}_{m-(k+1),1}(\tau) + \\
& \phantom{=} + \mathcal{N}^{\rm pd}_{k,1}(\tau) \, \mathcal{N}^{\rm pd}_{m-(k+1),2}(\tau) \, .
\end{cases}
\label{sol1_1_2}
\end{equation}
The solution of the system~(\ref{sol1_1}) is trivial and is given by 
\begin{equation}
\mathcal{N}^{\rm pd}_{0,j} (\tau) =- N_{{\rm i},j} \, \left [ \Exp{-\frac{D_j^2(\tau)}{2}} - 
\Exp{-\frac{D_j^2(1)}{2}}\right ] \, .
\label{sol0}
\end{equation}
Equation~(\ref{sol1_1_2}) can be solved by re-casting it into form~(\ref{prototype}) thus giving
\begin{equation}
\begin{cases}
\dot{\mathcal{N}}^{\rm pd}_{m,1}(\tau) &= \displaystyle
  -\sum_{k=0}^{m-1}
  \left(
    \mathcal{N}^{\rm bo}_{k,1}(\tau) + \mathcal{N}^{\rm pd}_{k,1}(\tau)
  \right)
  \left(
    \mathcal{N}^{\rm pd}_{m-(k+1),1}(\tau) + \xi_{m-(k+1)}
  \right) 
\\& \phantom{=} \qquad\qquad\qquad
  +\mathcal{N}^{\rm bo}_{k,2}(\tau) \, \mathcal{N}^{\rm pd}_{m-(k+1),1}(\tau)
  \\
& \qquad \qquad \qquad \qquad \qquad \qquad \qquad \qquad \qquad \qquad \qquad \qquad \qquad m \geq 1 \\
\mathcal{N}^{\rm pd}_{m,2}(\tau) &=\phantom{-}
  {\mathcal{N}}^{\rm pd}_{m,1}(\tau) + \xi_m
  \, ,
\end{cases}
\end{equation}
with solutions
\begin{equation}
\begin{cases}
{\mathcal{N}}^{\rm pd}_{m,1}(\tau) &=\displaystyle
  -\sum_{k=0}^{m-1}\Int{1}{\tau}{\tilde{\tau}} \bigg[
  \left(
    \mathcal{N}^{\rm bo}_{k,1}(\tilde{\tau})
    +\mathcal{N}^{\rm pd}_{k,1}(\tilde{\tau})
  \right)
  \left(
    \mathcal{N}^{\rm pd}_{m-(k+1),1}(\tilde{\tau}) + \xi_{m-(k+1)}
  \right) 
\\& \phantom{=} \qquad\qquad\qquad
  +\mathcal{N}^{\rm bo}_{k,2}(\tilde{\tau}) \,
    \mathcal{N}^{\rm pd}_{m-(k+1),1}(\tilde{\tau})
  \bigg]
  \\
& \qquad \qquad \qquad \qquad \qquad \qquad \qquad \qquad \qquad \qquad \qquad \qquad \qquad m \geq 1 \\
\mathcal{N}^{\rm pd}_{m,2}(\tau) &=\phantom{-}
  {\mathcal{N}}^{\rm pd}_{m,1}(\tau) + \xi_m
  \, ,
\end{cases}
\end{equation}

The functions $N^{\rm pd}_j(\tau)$ depend on the parameters $\xi_m$, which can generate a difference between $N^{\rm pd}_1(\tau)$ and $N^{\rm pd}_2(\tau)$ for $\tau=1$, given that $N^{\rm pd}_1(1)=0$. As the $N^{\rm pd}_j(\tau)$ represent the components related with the pseudo-diffusive effects, it is reasonable to set $\xi_m =0, m \geq 1$ as any intensity imbalance between the two beams is taken care of by the functions $N^{\rm bo}_j(\tau)$ and $\mathcal{N}^{\rm pd}_{0,j}(\tau)$.

If additional time dependencies are present, i.e. $\varepsilon=\varepsilon(\tau)$, extra care should be taken as a plain power expansion of $N^{\rm bo}_j(\tau), N^{\rm pd}_j(\tau)$ in $\varepsilon$ is not appropriate. Instead, we can write $\varepsilon$ as:
\begin{equation}
  \varepsilon(\tau) =
    \varepsilon_1 \; %\frac{\epsilon^*(1)}{\epsilon^*(\tau)}\frac{F(\tau)}{F(1)}\, ,
    \hat{\varepsilon}(\tau)
\end{equation}
where $\varepsilon_1=\varepsilon(1)$ its initial value. Note that
\begin{equation}
  \varepsilon_1 \ll 1\, , \qquad
  %\frac{\epsilon^*(1)}{\epsilon^*(\tau)}\frac{F(\tau)}{F(1)} \sim 1\, ,
   \hat{\varepsilon}(\tau)\sim 1\, ,
\end{equation}
implying we can expand $N^{\rm bo}_j(\tau), N^{\rm pd}_j(\tau)$ in function of the small (and constant) parameter $\varepsilon_1$:
\begin{equation}
 N^{\rm bo}_j(\tau) = \sum_{m=0}^{+\infty} \mathcal{N}^{\rm bo}_{m,j}(\tau) \, \varepsilon^m_1 \, , \qquad 
 N^{\rm pd}_j(\tau)= \sum_{m=0}^{+\infty} \mathcal{N}^{\rm pd}_{m,j}(\tau) \, \varepsilon^m_1 
\, , \qquad j=1,2 \, .
\label{sol11}
\end{equation}
As before, we replace these developments into Eq.~\eqref{int_main4} (now with $\varepsilon\to\varepsilon(\tau)$). This leaves Eq.~\eqref{sol1_1} unchanged, so the $\mathcal{O}(\varepsilon_1^0)$ solution is still valid as given by Eq.~\eqref{sol0}. For the higher order terms we can use the same steps as before, in order to finally get the full solution with time-dependent $\varepsilon(\tau)$:
\begin{equation}
\begin{cases}
{\mathcal{N}}^{\rm pd}_{m,1}(\tau) & = \displaystyle
  -\sum_{k=0}^{m-1} \Int{1}{\tau}{\tilde{\tau}}
    \hat{\varepsilon}(\tau) \bigg[
  \left(
    \mathcal{N}^{\rm bo}_{k,1}(\tilde{\tau})
    +\mathcal{N}^{\rm pd}_{k,1}(\tilde{\tau})
  \right) \times 
\\& \phantom{=}
  \times \left(
    \mathcal{N}^{\rm pd}_{m-(k+1),1}(\tilde{\tau}) + \xi_{m-(k+1)}
  \right) 
  +\mathcal{N}^{\rm bo}_{k,2}(\tilde{\tau}) \,
    \mathcal{N}^{\rm pd}_{m-(k+1),1}(\tilde{\tau})
  \bigg]
  \\
& \qquad \qquad \qquad \qquad \qquad \qquad \qquad \qquad \qquad \qquad \qquad \qquad \qquad m \geq 1 \\
\mathcal{N}^{\rm pd}_{m,2}(\tau) & =
  \phantom{-} {\mathcal{N}}^{\rm pd}_{m,1}(\tau) + \xi_m \, .
\end{cases}
\end{equation}

\begin{thebibliography}{99}
%
\bibitem{lumi_Part_II} M.~Giovannozzi, F.F.~Van~der~Veken, ``Description of the luminosity evolution for the CERN LHC including dynamic aperture effects. Part II: application to Run~1 data'', submitted to publication to Nucl. Instr. Meth. Phys. Res. A.
%
\bibitem{dynap1} M.~Giovannozzi, W.~Scandale, E.~Todesco, ``Prediction of long-term stability in large hadron colliders'', Part. Accel. {\bf 56} 195, 1996.
%
\bibitem{dynap2} M.~Giovannozzi, W.~Scandale, E.~Todesco, ``Dynamic aperture extrapolation in presence of tune modulation'', Phys. Rev. E {\bf 57} 3432, 1998.
%
\bibitem{loginvb-b} M.~Giovannozzi, E.~Laface, ``Investigations of Scaling Laws of Dynamic Aperture with Time for Numerical Simulations including Weak-Strong Beam-Beam Effects'', TUPPC086, IPAC12 proceedings, p.~1359, 2012.
%
\bibitem{lossesPRSTAB} M.~Giovannozzi, ``Proposed scaling law for intensity evolution in hadron storage rings based on dynamic aperture variation with time'', Phys. Rev. ST Accel. Beams {\bf 15} 024001, 2012.
%
\bibitem{DAexp_nekor} E.~Maclean, M.~Giovannozzi and R.~Appleby, ``Novel method to measure the extent of the stable phase space region of proton synchrotrons using Nekoroshev-like scaling law'', in preparation.
%
\bibitem{Lumi_fit} M.~Giovannozzi, C.~Yu, ``Proposal of an Inverse Logarithm Scaling Law for the Luminosity Evolution'', TUPPC078, IPAC12 proceedings, p.~1353, 2012.
%
\bibitem{IPAC14} M.~Giovannozzi, ``Simple models describing the time-evolution of luminosity in hadron colliders'', TUPRO009, IPAC14 proceedings, p.~1017, 2014.
%
\bibitem{inel1} The ATLAS and CMS Collaborations, ``Expected pile-up values at HL-LHC'', ATL-UPGRADE-PUB-2013-014, 2013.
%
\bibitem{inel2} D.~Contardo, private communication, 3 December 2014.
%
\bibitem{bruce} R. Bruce, J. M. Jowett, M. Blaskiewicz, W. Fischer, ``Time evolution of the luminosity of colliding heavy-ion beams in BNL Relativistic Heavy Ion Collider and CERN Large Hadron Collider'', Phys. Rev. ST Accel. Beams {\bf 13}, 091001, 2010.
%
\bibitem{fnal1} E.~McCrory, V.~Shiltsev, A.~J. Slaughter, A.~Xiao, ``Fitting the luminosity decay in the Tevatron'', TPAP036, PAC 2005 proceedings, p.~2434, 2005.
%
\bibitem{fnal2} V.~Shiltsev, E.~McCrory, ``Characterizing luminosity evolution in the Tevatron'', TPAP038, PAC 2005 proceedings, p.~2536, 2005.
%
\bibitem{FZ} M. Benedikt, D. Schulte, F. Zimmermann, ``Optimizing integrated luminosity of future hadron colliders'', Phys. Rev. ST Accel. Beams {\bf 18}, 101002 (2015).
%
\bibitem{riccati} S.~Bittanti, A.~J.~Laub, J.~C.~Willems (eds.), {\sl The Riccati equation}, Berlin Springer-Verlag, 1991. 
%
\bibitem{KAM} C.~L.~Siegel and J.~Moser, {\sl Lectures in celestial mechanics}, Berlin Springer-Verlag, 1971.
%
\bibitem{nekhor} N.~Nekhoroshev, ``An exponential estimate of the time of stability of nearly-integrable Hamiltonian systems'', Russ. Math. Surv. {\bf 32} 1, 1977.
%
\bibitem{gtnekhor} A.~Bazzani, S.~Marmi, G.~Turchetti, ``Nekhoroshev estimate for isochronous non resonant symplectic maps '', Cel. Mech. {\bf 47} 333, 1990.
%
\bibitem{stoch1} M. Blaskiewicz and J. M. Brennan, ``Bunched beam stochastic cooling in a collider'', Phys. Rev. ST Accel. Beams {\bf 10}, 061001, 2007.
%
\bibitem{stoch2} M. Blaskiewicz, J. M. Brennan, and F. Severino, ``Operational Stochastic Cooling in the Relativistic Heavy-Ion Collider'', Phys. Rev. Lett. {\bf 100}, 174802, 2008.
%
\bibitem{stoch3} M. Blaskiewicz, J. M. Brennan, and K. Mernick, ``Three-Dimensional Stochastic Cooling in the Relativistic Heavy Ion Collider'', Phys. Rev. Lett. {\bf 105}, 094801, 2010. 
%
\bibitem{hilumi} L.~Rossi, ``LHC Upgrade Plans: Options and Strategy'', TUYA02, IPAC11 proceedings, p.~908, 2011.
%
\bibitem{PDR} G.~Apollinari, I.~B\'ejar Alonso, O.~Br\"uning, M.~Lamont, L.~Rossi, ``High-Luminosity Large Hadron Collider (HL-LHC): Preliminary Design Report'', CERN-2015-005, 2015.
%
\bibitem{cc} R.~Calaga, J.-P.~Koutchouk, R.~Tom\'as Garcia, J.~Tuckmantel, F.~Zimmermann, ``LHC Crab-Cavity Aspects and Strategy'', TUOAMH02, IPAC10 proceedings, p.~1240, 2010.
%
\bibitem{handbook} A.~W.~Chao, M.~Tigner, {\sl Handbook of Accelerator Physics and Engineering}, World Scientific, 1998.
%
\bibitem{bryant} P.~J.~Bryant, K.~Johnsen, {\sl The Principles of Circular Accelerators and Storage Rings}, Cambridge University Press, 1993.
%
\bibitem{Courant-Snyder} E.~Courant and H.~Snyder, ``Theory of the Alternating-Gradient Synchrotron'', Ann. Phys. {\bf 3}, 1, 1958.
%
\bibitem{LHCDR} O.~S.~Br\"uning, P.~Collier, Ph.~Lebrun, S.~Myers, R.~Ostojic, J.~Poole, P.~Proudlock, ``LHC Design Report, v.1: the LHC Main Ring'', CERN-2004-003-V-1, 2004.
%
\end{thebibliography}
\end{document}